\documentclass[reprint,amsmath,amssymb,aps,pra]{revtex4-1}
\usepackage{graphicx}
\usepackage{braket}

\begin{document}
\title{Quantum Purification for Amplitude Damping Noise}
\author{Kai Wang\textsuperscript{1}}
\author{Zhen-Yang Peng\textsuperscript{2}}
\email{pengzhenyang@sjtu.edu.cn}
\affiliation{\textsuperscript{1}\mbox{Interdisciplinary Artificial Intelligence Research Institute, Wuhan College, Wuhan 430212,China} 
\\
\textsuperscript{2}Wilczek Quantum Center, School of Physics and Astronomy, Shanghai Jiao Tong University, 200240, China}
\date{\today}
         
\begin{abstract}
    Noise poses a fundamental challenge to quantum information processing, 
    with amplitude-damping (AD) noise being particularly detrimental. 
    Preserving high-fidelity quantum systems therefore 
    relies critically on effective error correction and purification methods. 
    In this work, 
    we introduce a novel approach for mitigating AD noise 
    that can be applied to both state and channel purification. 
    Our method achieves a substantial enhancement in the fidelity 
    of affected states or channels
    while maintaining a low resource overhead, 
    requiring only one or two ancilla qubits 
    in combination with two Clifford gates,
    and exhibits a relatively high success probability. 
    This approach provides a practical and scalable framework 
    for addressing AD noise in realistic quantum systems. 
\end{abstract}

\maketitle

\section{\label{intro}Introduction}
    Quantum information processing 
    is fundamentally limited by noise, 
    which degrades the fidelity of quantum states and operations. 
    Among various noise types, 
    amplitude-damping (AD) noise is particularly detrimental,
    as it models energy loss processes that occur in most physical implementations \cite{book_nielsen_chuang_quantum_2010,preskill_2015lecture,AD_Kashif2024InvestigatingTE}.
    To achieve reliable quantum computation and communication, 
    both quantum error correction and quantum state purification are essential. 

    Quantum error correction \cite{book_lidar2013quantum} 
    actively protects quantum information mostly
    by encoding logical qubits into redundant Hilbert spaces
    and detecting or correcting errors 
    without directly measuring the encoded information \cite{Shor_1995,code_steane_multiple-particle_1996,code_5_qubits,Stabilizer_Gottesman1997CA,25_years_of_quantum_error_correction,surface_code_2025,topological_code,Topological-subsystem-codes,color_code,color_code_2,Demonstration_Steane_code,PhysRevLett.132.250602}. 
    There are also channel adapted codes designed specifically for AD noise \cite{ad_code_1997,ad_code_2008,adcode-2010,adcode-2011,ad_code_2013,ad_code_2014,adcode-2020}. 
    In contrast, purification is a probabilistic procedure 
    that increases the fidelity of a quantum state \cite{Purify_state,Purify-state-2001,purify-state-2017,purify-state-2020} or channel \cite{purifychannel2003,purifychannel2022,purify_filter_channel2024,purify_sym_channel2025,purifychannel2025}, 
    typically by post-selecting measurement outcomes or averaging over multiple noisy copies. 
    The latter approach is often referred to as virtual purification, 
    as it effectively improves only the fidelity of expectation values 
    rather than producing an actual purified state \cite{virtual_2022,virtual_2024,virtual_dis_2024,purify_sym_channel2025,purifychannel2025}.

    Purification is particularly useful in scenarios 
    where error correction is costly or impractical, 
    such as near-term quantum devices with limited qubit resources \cite{limit_qubit}, 
    or when one seeks to distill high-fidelity states from multiple noisy copies 
    for tasks including entanglement distribution \cite{entanglement-distribution_2001,entanglement-distribution_2019,entanglement-distribution_2024}, 
    quantum communication \cite{quantum-communication-2019}, 
    and certain quantum algorithms \cite{Quantum-algorithms-2016,Quantum-algorithms2021}.
    It is important to note that 
    purification can be applied exclusively to pure states 
    and cannot be directly used for mixed states,
    whereas error correction can handle both scenarios,
    making error correction more broadly applicable. 
    However, according to the Choi-Jamio{\l}kowski isomorphism \cite{JAMIOLKOWSKI-1972275,CHOI-state}, 
    any quantum channel can be mathematically represented as a pure state. 
    Combined with the complex circuits and large amounts of ancillas 
    required for error correction, 
    leading purification a particularly efficient and practical approach 
    for mitigating noise in both quantum states and high-dimensional channels.

    In this work, 
    we focus on purification for AD noise, 
    proposing a method applicable to both state and channel purification 
    that enhances the fidelity of affected quantum systems 
    using only a single or at most two ancilla qubits and 
    two Clifford gates: the Hadamard and controlled-Z gates. 
    Unlike previous purification methods 
    that require multiple copies of the states to be protected, 
    our circuit operates on a single noisy state, 
    reducing the circuit dimension and 
    thereby decreasing the likelihood of additional noise, 
    while circumventing the limitations imposed by the no-cloning theorem \cite{no-cloning}. 
    Furthermore, the circuit can also be employed to probe the effects of AD noise. 
    Our approach provides a practical and efficient strategy 
    for mitigating dominant decoherence mechanisms in realistic quantum systems.
               
    This paper is organized as follows. 
    We begin in Section~\ref{sec2} with the necessary background information 
    on the AD noise channel and a brief introduction to purification, 
    establishing the notation and terminology used throughout the paper.
    In Section~\ref{sec3}, 
    we introduce our setup for state purification and 
    demonstrate how it can also be employed to probe AD noise. 
    In Section~\ref{sec4}, 
    we extend the method to multi-qubit purification 
    for noise with equal probabilities on each qubit, 
    which also serves as a strategy for channel purification under AD noise. 
    Section~\ref{sec5} presents numerical results 
    evaluating both the fidelity of our methods, 
    as well as the corresponding success probabilities,
    validating their effectiveness and efficiency. 
    Finally, 
    we conclude and summarize our work in Section~\ref{sec6}.

\section{\label{sec2}Preliminary}
    \subsection{AD noise}
        AD noise describes energy relaxation from the excited state $\ket{1}$ 
        to the ground state $\ket{0}$ 
        due to system--environment coupling. 
        In general, 
        any quantum process can be modeled as a quantum channel $\Phi$,
        also known as a completely positive trace-preserving (CPTP) map \cite{CHOI-state}. 
        Such a map can be expressed in terms of a set of Kraus operators $\{K_i\}$ as
        \begin{equation}
           \Phi(\rho) = \sum_i K_i \rho K_i^\dagger, 
        \end{equation}
        where $\sum_i K_i^\dagger K_i = I$ guarantees 
        complete positivity and trace preservation \cite{kraus-operator}. 
        For the AD channel $\mathcal{E}_{\mathrm{AD}}$, 
        the Kraus operators are given by
        \begin{equation}
        \label{AD_E0_E1}
        E_0 = 
        \begin{pmatrix}
        1 & 0 \\
        0 & \sqrt{1-\gamma}
        \end{pmatrix},
        \qquad
        E_1 = 
        \begin{pmatrix}
        0 & \sqrt{\gamma} \\
        0 & 0
        \end{pmatrix},
        \end{equation}
        where $\gamma \in [0,1]$ is the damping probability. 
        Physically, 
        $\gamma$ represents the probability that 
        an excitation decays during the noise interval.

    \subsection{Purification}
        Suppose qubits are initially prepared in an unknown pure state $\ket{\phi}$. 
        After undergoing noise, 
        which can be described by a quantum channel $C_E$ 
        with Kraus operators $\{K_i\}$, 
        the state evolves into
        \begin{equation}
            \rho_E=C_E(\ket{\phi}\bra{\phi}) = 
            \sum_i K_i \ket{\phi}\bra{\phi} K_i^\dagger.
        \end{equation}
        The central goal of purification is 
        to recover the maximum possible number of qubits 
        in a state that is “as close as possible” to the original unknown state $\ket{\phi}$.
        
        The quality of purification is typically evaluated in terms of 
        both the fidelity~\cite{book_nielsen_chuang_quantum_2010} 
        between the original state $\ket{\phi}$ and 
        the purified state $\rho$,  
        \begin{equation}
        \label{eq:fid}
            f(\ket{\phi}\bra{\phi}, \rho) = f (\ket{\phi}\bra{\phi},
            \mathcal{C}_{\mathrm{purify}} \circ C_E(\ket{\phi}\bra{\phi}),
        \end{equation}  
        where we use $\mathcal{C}_{\mathrm{purify}}$ to 
        describe the overall purification procedure, 
        and the probability $p$ of successfully obtaining the desired outcome.  
        These two criteria are often in competition: 
        higher fidelity generally comes at the cost of a lower success probability, and vice versa.  

        In practice, 
        purification can be realized either through post-selection, 
        where only favorable measurement outcomes are kept  
        while the rest states, 
        severely affected by noise, are discarded, 
        or through virtual purification,  
        which consists of averaging over multiple noisy realizations to 
        mitigate the average effect of the noise,
        thereby improving the fidelity of expectation values. 
        Although virtual purification does not produce an actual high-fidelity quantum state, 
        it is deterministic—no runs are discarded—and is therefore resource-efficient.  
        This makes it particularly suitable for near-term experiments with limited samples, 
        where full-fledged error correction is not yet feasible.  
        By contrast, 
        purification via post-selection has the key advantage 
        of producing an actual purified quantum state, 
        which can be directly reused in subsequent computational or communication tasks.  
        Moreover, 
        post-selection can in principle achieve higher fidelity 
        with respect to the original state $\ket{\phi}$, 
        since unfavorable noisy realizations are explicitly filtered out,
        albeit at the potential cost of reduced success probability. 
        In this work, 
        we achieve purification via post-selection.

\section{\label{sec3}SETUP FOR STATE PURIFICATION}
\begin{figure}[!htbp]
        \centering\includegraphics[width=0.45\textwidth]{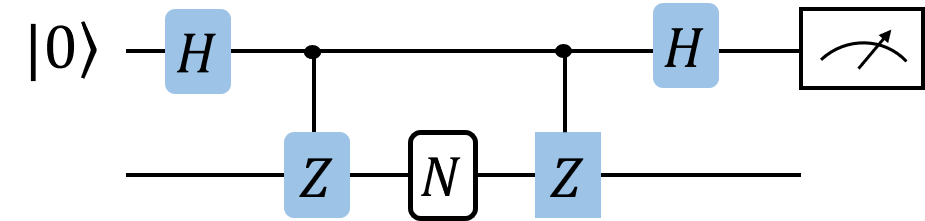}
        \caption{(Color online) Circuit for state purification under AD noise, 
        where $N$ represents the AD noise channel 
        and the first qubit serves as an ancilla.}
        \label{fig.circuit}
    \end{figure}

    Our circuit for state purification under AD noise is shown in Fig.~\ref{fig.circuit},
    where $N$ represents the AD noise channel.  
    The first qubit serves as an ancilla: 
    by measuring it and post-selecting on the desired outcome, 
    the noise-induced errors on the second qubit 
    can be detected and effectively filtered out, 
    resulting in a purified state.  

    To illustrate the operation of the circuit, 
    consider an arbitrary input state of the second qubit $\ket{\psi}$.
    The output state of both qubits before measurement then is 
    \begin{equation}
        \ket{\xi}= 
        \ket{+} \otimes \sum_i \sqrt{p_i} E_i \ket{\psi}
         + \ket{-} \otimes ( Z \sum_i \sqrt{p_i} E_i  Z ) \ket{\psi}, 
    \end{equation}
    where 
    \begin{equation}
    \label{eq:prob}
        p_i = \langle \psi \vert E_i^\dagger E_i \vert \psi \rangle
    \end{equation}  
    denotes the probability that the noise channel acts through the Kraus operator $E_i$, 
    whose explicit matrices are given in Eq.~\eqref{AD_E0_E1}.
    Consequently, 
    depending on the measurement outcome of the ancilla, 
    the second qubit collapses to one of the corresponding conditional states:
    \begin{align}
        & \label{eq:collapse0} 
        \ket{0}:
        \sum_i \sqrt{p_i} E_i \ket{\psi} + ( Z \sum_i \sqrt{p_i} E_i  Z ) \ket{\psi},  \\
        & \label{eq:collapse1}
        \ket{1}: \sum_i \sqrt{p_i} E_i \ket{\psi} -  ( Z \sum_i \sqrt{p_i} E_i  Z ) \ket{\psi}.
    \end{align}
    Noted for AD noise, its Kraus operators satisfy:
    \begin{equation}
        [E_0, Z]=0, \quad \{ E_1, Z \} =0,
    \end{equation}
    which simply Eqs.~\eqref{eq:collapse0} and \eqref{eq:collapse1} into:
    \begin{align}
        & \label{collapse0'} 
        \ket{0}:
          E_0 \ket{\psi},  \\
        & \label{collapse1'}
        \ket{1}:  E_1 \ket{\psi},
    \end{align}
    with the corresponding measurement probabilities $p_0$ and $p_1$, respectively.  
    Therefore, by measuring the ancilla and post-selecting on the $\ket{0}$ outcome, 
    the circuit shown in Fig.~\ref{fig.circuit} effectively 
    realizes state purification under AD noise.
    A straightforward calculation shows that for any state to be purified,
    \(\ket{\psi} = \alpha \ket{0} + \beta \ket{1}\), 
    the probability of obtaining the ancilla measurement outcome \(\ket{0}\) is
    \begin{equation}
    \label{pro_state}
        p_0 = \alpha^2 + (1-\gamma)\beta^2,
    \end{equation}
    and the corresponding purified state of the second qubit, 
    in its unnormalized form, is 
    \begin{equation}
        \ket{\psi'} = \alpha \ket{0} + \sqrt{1-\gamma}\, \beta \ket{1}.
    \end{equation}
    The fidelity of this purification is then given by
    \begin{equation}
        f(\ket{\psi},\ket{\psi'}) = |\langle \psi | \psi' \rangle|^2 
        = \frac{(\alpha^2 + \sqrt{1-\gamma}\, \beta^2)^2}{\alpha^2 + (1-\gamma) \beta^2}.
    \end{equation}

    Additionally, 
    there exist some error-correction schemes for AD noise 
    that require an estimation of the damping parameter~$\gamma$
    in advance~\cite{AD_parameter,AD_parameter_PhysRevLett.95.090501}. 
    The circuit in Fig.~\ref{fig.circuit} can also be used to 
    estimate the strength of AD noise, 
    since the measurement outcomes $\ket{i=0,1}$ occur with probabilities 
    \( 
        p_i 
    \)
    as given in Eq.~\eqref{eq:prob}. 
    Explicitly, 
    for $i=0,1$ we have
    \begin{equation}
        p_0 = \bra{\psi} 
        \begin{pmatrix}
        1 & 0 \\
        0 &  1-\gamma 
        \end{pmatrix}
        \ket{\psi},
        \qquad
        p_1 = \bra{\psi} 
        \begin{pmatrix}
        0 & 0 \\
        0 &  \gamma
        \end{pmatrix}
        \ket{\psi}.
    \end{equation}  
    The parameter $\gamma$ can then be directly estimated
    by preparing the input state of the second qubit as $\ket{1}$. 
    In this case, the probability of obtaining the outcome $\ket{1}$ is 
    \begin{equation}
      p_1 = \bra{1} E_1^\dagger E_1 \ket{1} = \gamma,
    \end{equation}
    providing a straightforward probe of the AD noise strength.  
    Moreover, 
    once $\gamma$ is estimated, 
    one can further purify the second qubit in our setup 
    by applying the non-unitary operator
    \begin{equation}
        \begin{pmatrix}
            1 & 0 \\
            0 & 1/\sqrt{1-\gamma}
        \end{pmatrix},
    \end{equation}
    which compensates for the attenuation 
    introduced by the no-jump operator $E_0$, 
    thereby restoring the relative weight of the excited state $\ket{1}$.
 
\section{\label{sec4}SETUP FOR CHANNEL PURIFICATION}
\begin{figure}[! htbp]
        \centering\includegraphics[width=0.45\textwidth]{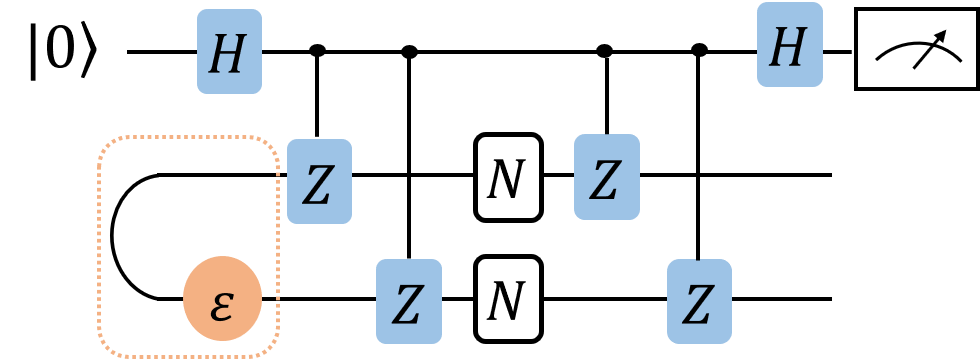}
        \caption{(Color online) Circuit for channel purification under AD noise, 
        where $N$ represents the AD noise channel, 
        $\varepsilon$ denotes the channel to be purified
        and the first qubit serves as an ancilla.}
        \label{fig.circuit_for_channel}
    \end{figure}

    We now extend our method to channel purification under AD noise.
    As established by the Choi--Jamio{\l}kowski isomorphism, 
    every quantum channel \(\Phi\) can be uniquely represented by 
    a corresponding Choi state, 
    \begin{equation}
        \rho_{\text{Choi}} = (I \otimes \Phi)\,\ket{\phi_{\text{Bell}}}\bra{\phi_{\text{Bell}}},
    \end{equation}
    where 
    \begin{equation}
      \ket{\phi_{\text{Bell}}} = \frac{1}{\sqrt{2}}(\ket{00}+\ket{11})  
    \end{equation}
    denotes the maximally entangled Bell state~\cite{bell_state}. 
    Thus, by preparing the Choi state associated with the channel to be purified, 
    one can employ analogous purification strategies 
    to mitigate AD noise in a quantum channel. 
    The corresponding circuit is shown in Fig.~\ref{fig.circuit_for_channel}, 
    where \(N\) denotes the AD noise channel, 
    which may act on both input qubits. 
    The region enclosed by the dashed line 
    represents the input Choi state of channel \(\varepsilon\), 
    and the first qubit serves as an ancilla.  

    Consider an arbitrary input channel $\varepsilon$,
    with its corresponding Choi state 
    \begin{equation}
    \begin{aligned}
       \rho_\text{Choi} 
        & = \sum_{m,n} \ket{m}\bra{n} \otimes \varepsilon(\ket{m}\bra{n})\\
        &  =\sum_{m,n} \ket{m}\bra{n} \otimes \sum_\alpha K_\alpha \ket{m}\bra{n} K_\alpha^\dagger,
    \end{aligned}
    \end{equation}
    where \(\{K_\alpha\}\) denote the Kraus operators of the channel \(\varepsilon\),
    and $m,n=0,1$.
    The output state before measurement can then be expressed as
    \begin{equation}
    \begin{aligned}
        \ket{\xi} = & \sum_{m} \sum_{\alpha} \sum_{i,j} \Big[ 
            \ket{+} \otimes 
            \big( \sqrt{p_{im}}\, E_i \ket{m} \otimes   \sqrt{p_{j\alpha m} \,  q_{\alpha m}}\, \cdot
            \\ & E_j K_\alpha \ket{m} \big) 
             + \ket{-} \otimes 
            \big( \sqrt{p_{im}}\, Z E_i Z \ket{m} \otimes \\
            & \sqrt{ p_{j\alpha m} \,   q_{\alpha m} }\, Z E_j Z K_\alpha \ket{m} \big) 
        \Big],
    \end{aligned}
    \end{equation}
    in terms of the probabilities
    \begin{align}
        p_{im} &= \bra{m}\, E_i^\dagger E_i \,\ket{m}, \label{eq:pim} \\
        q_{\alpha m} &= \bra{m}\, K_\alpha^\dagger K_\alpha \,\ket{m}, \label{eq:qalpham} \\
        p_{j\alpha m} &= \bra{m}\, K_\alpha^\dagger E_j^\dagger E_j K_\alpha \,\ket{m}, \label{eq:pjalpham}
    \end{align}
    which represent the probabilities associated with 
    the Kraus operators \(E_i\), \(K_\alpha\), 
    and their combined action on the basis state \(\ket{m}\), respectively.
    For AD noise, a straightforward calculation gives
    \begin{equation}
        p_{00} = 1, \quad p_{01} = 1-\gamma, \quad p_{10} = 0, \quad p_{11} = \gamma.
    \end{equation}
    Depending on the ancilla measurement outcome, 
    the second and third qubits collapse into the corresponding post-measurement states:
    \begin{align}
        \ket{0}: 
        & \sum_{m} \sum_{\alpha} \sum_{i,j}
        \Big( \sqrt{p_{im}}\, E_i \ket{m} \otimes 
              \sqrt{p_{j\alpha m}\, q_{\alpha m}}\, E_j K_\alpha \ket{m} \notag \\
        & \quad + \sqrt{p_{im}}\, Z E_i Z \ket{m} \otimes 
              \sqrt{p_{j\alpha m}\, q_{\alpha m}}\, Z E_j Z K_\alpha \ket{m} \Big), \\
        \ket{1}:  
        & \sum_{m} \sum_{\alpha} \sum_{i,j}
        \Big( \sqrt{p_{im}}\, E_i \ket{m} \otimes 
              \sqrt{p_{j\alpha m}\, q_{\alpha m}}\, E_j K_\alpha \ket{m} \notag \\
        & \quad - \sqrt{p_{im}}\, Z E_i Z \ket{m} \otimes 
              \sqrt{p_{j\alpha m}\, q_{\alpha m}}\, Z E_j Z K_\alpha \ket{m} \Big),
    \end{align}
    which can be further simplified under AD noise yielding
    \begin{align} 
        \ket{0}: 
        & \sum_{m} \sum_{\alpha} \Big[ 
        \sqrt{p_{0m} p_{0\alpha m}\, q_{\alpha m}}\,
        \Big( E_0 \otimes E_0 \Big)
        \Big(   \ket{m} \otimes 
              K_\alpha \ket{m} \Big) \notag \\ 
         &     + \sqrt{\gamma p_{1\alpha 1}\, q_{\alpha 1}}\,
         \Big( E_1 \otimes E_1 \Big)
        \Big(   \ket{1} \otimes 
                K_\alpha \ket{1} \Big) \Big] \label{eq:col0_channel} ,\\ 
        \ket{1}:  
        & \sum_{m} \sum_{\alpha}   \Big[  \sqrt{p_{0m} p_{1\alpha m}\,q_{\alpha m}}
        \Big( E_0 \otimes E_1 \Big)
        \Big(   \ket{m} \otimes 
             K_\alpha \ket{m} \Big) \notag \\ 
        &      + \sqrt{\gamma p_{0\alpha 1}\, q_{\alpha 1}}\,
        \Big( E_1 \otimes E_0 \Big)
        \Big(   \ket{1} \otimes 
                K_\alpha \ket{1} \Big) \Big].
    \end{align}
    It is worth noting that error correction or purification is meaningful 
    only when the noise probability is sufficiently small,
    i.e., in this case, the parameter $\gamma$ is relatively small.
    In this regime, the occurrence probability of 
    $p(E_1 \otimes E_1 ) \propto \gamma^2 $ is strongly suppressed,
    so that the first term in Eq.~\eqref{eq:col0_channel} becomes dominant.
    This suggests that by measuring the ancilla  
    and post-selecting on the $\ket{0}$ outcome, 
    the circuit shown in Fig.~\ref{fig.circuit_for_channel} 
    can realize channel purification under AD noise.

\begin{figure}[! htbp]
        \centering\includegraphics[width=0.48\textwidth]{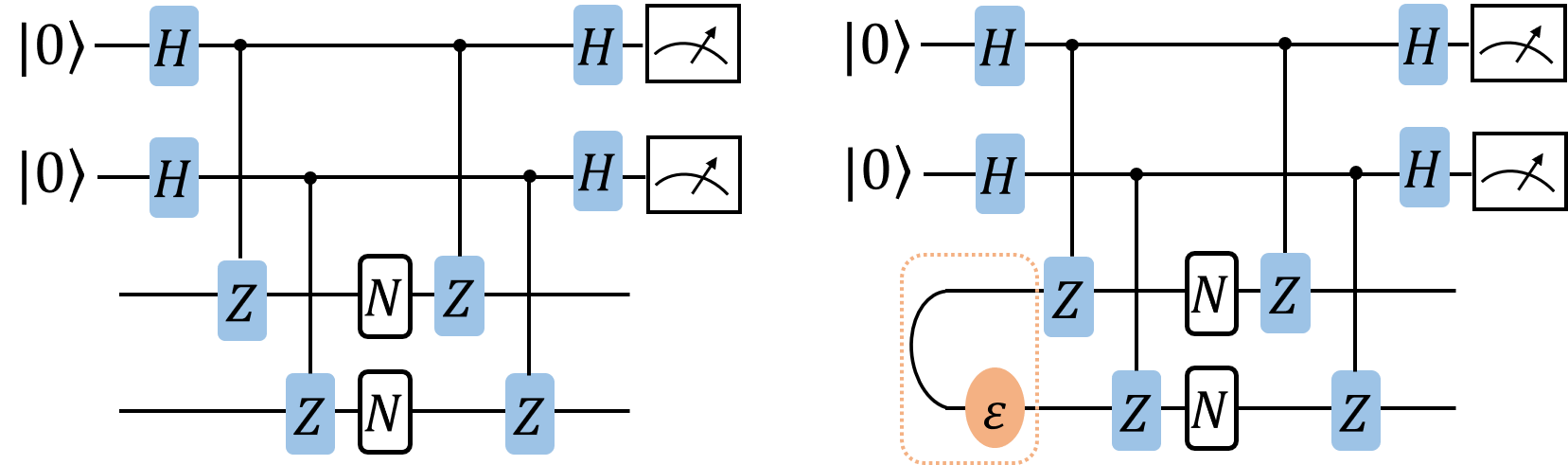}
        \caption{(Color online) Circuit for two-qubit state purification (left) 
        and channel purification (right) under AD noise using two ancillas, 
        where $N$ represents the AD noise channel and the first two qubits 
        serve as ancillas.}
        \label{fig.circuit_for_two_qubit}
    \end{figure}

    We also note by using two ancillas and the method illustrated in Fig.~\ref{fig.circuit}, 
    one can achieve purification of a two-qubit state, 
    as shown in Fig.~\ref{fig.circuit_for_two_qubit} (left).  
    Since a two-qubit state and its corresponding Choi state are interchangeable, 
    the same setup can also be employed to purify a quantum channel, 
    as illustrated in Fig.~\ref{fig.circuit_for_two_qubit} (right). 

    However, when ancilla resources are limited, 
    the circuit depicted in Fig.~\ref{fig.circuit_for_channel} 
    is more resource-efficient and practical.  
    At the same time, 
    this circuit can also be adapted for two-qubit state purification 
    by replacing the Choi state with the input two-qubit state to be purified.

    For the circuit shown in Fig.~\ref{fig.circuit_for_two_qubit}, 
    the measurement outcomes of the first two ancilla qubits 
    determine the post-measurement state of the last two qubits. 
    As expected, 
    for an input state $\ket{\psi}$ of the last two qubits, 
    the system collapses into the following states 
    depending on the measurement outcome:
    \begin{equation}
        \begin{aligned}
            \ket{00} &:\; E_0 \otimes E_0 \ket{\psi}, \\
            \ket{01} &:\; E_0 \otimes E_1 \ket{\psi}, \\
            \ket{10} &:\; E_1 \otimes E_0 \ket{\psi}, \\
            \ket{11} &:\; E_1 \otimes E_1 \ket{\psi}.
        \end{aligned}
    \end{equation}

\section{\label{sec5}Numerical calculation}
    To demonstrate the validity of our approach, 
    we numerically calculate the fidelities using Eq.~\eqref{eq:fid} 
    as well as the probabilities of obtaining the desired measurement outcomes
    for different AD parameters $\gamma$.  
    The circuit can be implemented directly or 
    simplified beforehand using the ZX-calculus~\cite{ZX-calculus}.

\begin{figure}[! htbp]
        \centering\includegraphics[width=0.45\textwidth]{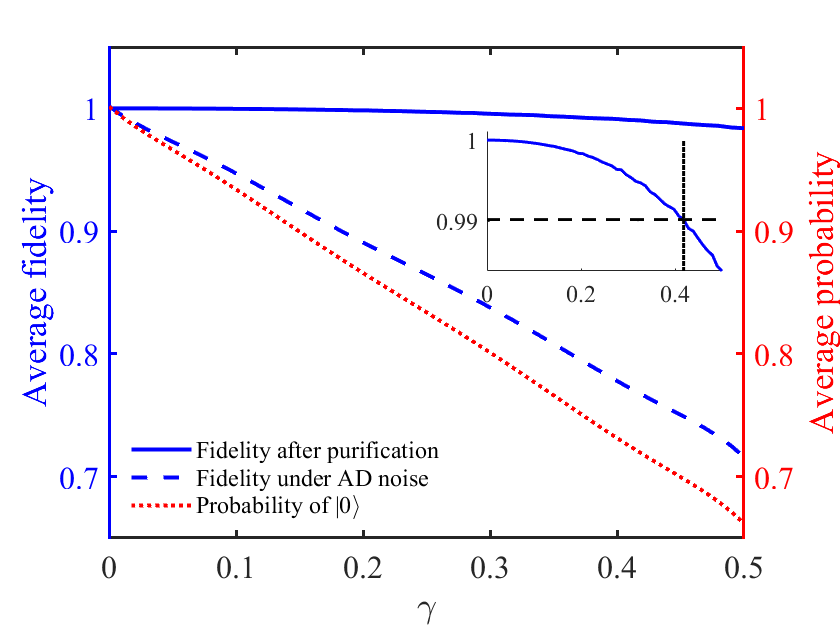}
        \caption{(Color online) Fidelity and probability as a function of the AD parameter $\gamma$. 
        The blue dashed curve depicts the average fidelity of the unpurified state under AD noise, 
        the blue solid curve corresponds to the average fidelity after purification, 
        and the red dotted curve represents the average probability of obtaining 
        the ancilla measurement outcome $\ket{0}$.}
        \label{fig.num_for_state}
    \end{figure}

\begin{figure}[! htbp]
        \centering\includegraphics[width=0.45\textwidth]{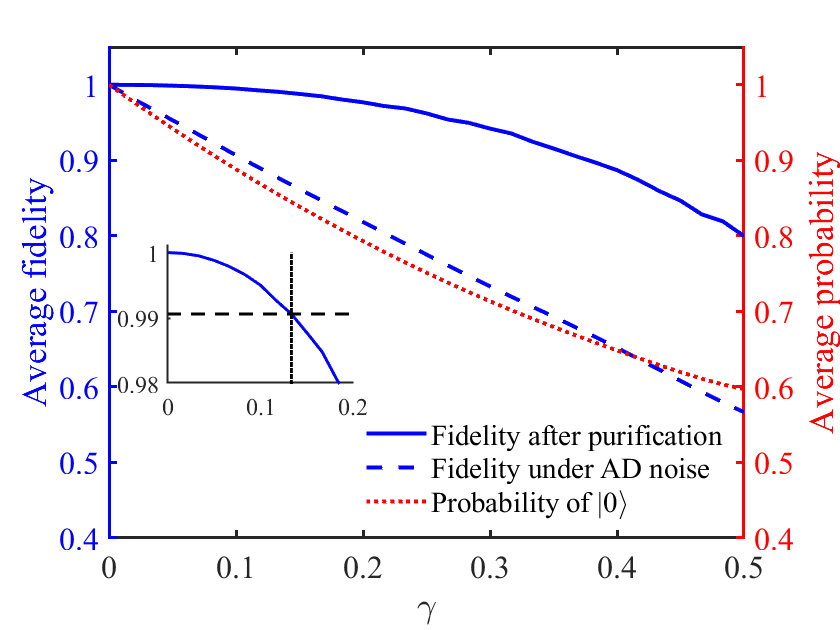}
        \caption{(Color online) Fidelity and probability as a function of the AD parameter $\gamma$. 
        The blue dashed curve depicts the average fidelity of the unpurified channel under AD noise, 
        the blue solid curve corresponds to the average fidelity after purification, 
        and the red dotted curve represents the average probability of obtaining 
        the ancilla measurement outcome $\ket{0}$.}
        \label{fig.num_for_channel}
    \end{figure}

\begin{figure}[! htbp]
        \centering\includegraphics[width=0.45\textwidth]{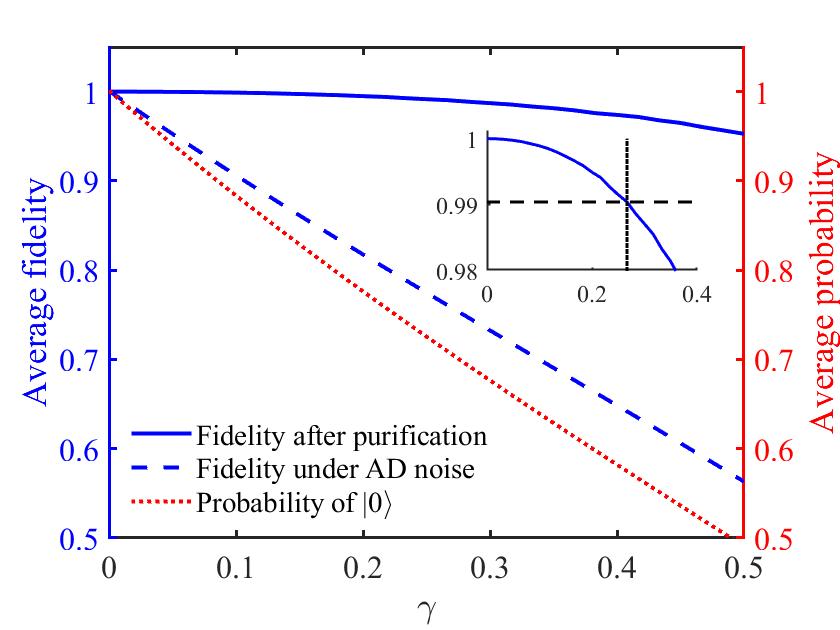}
        \caption{(Color online) Fidelity and probability as a function of the AD parameter $\gamma$. 
        The blue dashed curve depicts the average fidelity of the unpurified state under AD noise, 
        the blue solid curve corresponds to the average fidelity after purification, 
        and the red dotted curve represents the average probability of obtaining the ancillas measurement outcome $\ket{00}$.}
        \label{fig.num_for_two_qubit}
    \end{figure}

    The results are shown in Fig.~\ref{fig.num_for_state} for single-qubit state purification 
    and in Fig.~\ref{fig.num_for_channel} for channel purification, 
    both using one ancilla, 
    and in Fig.~\ref{fig.num_for_two_qubit} for two-qubit state purification using two ancillas.

    For all figures, 
    the input states or channels are randomly sampled 1000 times 
    for each chosen value of $\gamma$.
    The blue dashed curves depict the average fidelity 
    of the unpurified states under AD noise, 
    while the blue solid curves 
    correspond to the average fidelity after purification, 
    i.e., conditioned on obtaining the ancilla measurement outcome 
    $\ket{0}$ in the first two figures, 
    and $\ket{00}$ in the last figure.  
    The red dotted curves indicate the average probability 
    of obtaining the corresponding favorable ancilla measurement outcome.

    We highlight the critical points in each figure 
    where the fidelity after purification drops below $0.99$.  
    From the locations of these points, 
    it is evident that the single-qubit purification method is highly competitive, 
    achieving very high fidelity- 
    remaining above $0.99$ even for a relatively strong noise parameter, $\gamma = 0.4184$-  
    while simultaneously maintaining a reasonably high success probability, 
    with more than $71.94\%$ of runs yielding the desired measurement outcome.

    When AD noise acts on two qubits, 
    fidelities above $0.99$ are attainable 
    only when the noise strength is sufficiently low, 
    specifically for $\gamma < 0.1333$ when using a single ancilla 
    and $\gamma < 0.2667$ when using two ancilla qubits.  
    The corresponding success probabilities at these points 
    are $85.47\%$ and $70.90\%$, respectively.  
    By introducing one additional ancilla qubit, 
    the purification fidelity can be significantly enhanced, 
    remaining above $0.95$ even for a relatively strong noise parameter, $\gamma = 0.5$.  
    However, 
    this improvement comes at the cost of a decreased probability 
    of obtaining the favorable ancilla measurement outcome, 
    highlighting the inherent trade-off 
    between purification fidelity 
    and operational efficiency in the protocol.
 
    At the same time, 
    it is straightforward to observe from Fig.~\ref{fig.num_for_state} 
    that the probability of obtaining the ancilla measurement outcome $\ket{0}$ is 
    linearly anti-correlated with the AD parameter $\gamma$, 
    in excellent agreement with our analytical result given in Eq.~\eqref{pro_state}.  
    This behavior is also observed in the two-ancilla purification setup, 
    as both configurations follow the same underlying purification logic.  
    It is worth emphasizing that all success probabilities remain above $0.5$, 
    indicating that the proposed purification protocol is not only effective in enhancing fidelity 
    but also practical for experimental implementation.
    
    For two-qubit or channel purification, 
    the circuits shown in Fig.~\ref{fig.circuit_for_channel} 
    and Fig.~\ref{fig.circuit_for_two_qubit} 
    are essentially interchangeable.  
    However, 
    when the AD noise has a small damping parameter, 
    it is more resource-efficient to perform purification 
    using only a single ancilla.  
    This approach not only ensures high fidelity 
    but also offers greater operational efficiency 
    compared to the two-ancilla setup, 
    making it particularly suitable for low-noise scenarios.
    
\section{\label{sec6}Conclusion}
    In conclusion, 
    we have proposed a method for AD purification 
    that requires at most two ancilla qubits 
    and only two types of Clifford gates, 
    namely the Hadamard and CZ gates. 
    The method can be applied to both state and channel purification. 
    Unlike previous purification strategies
    that require multiple copies of the states to be purified, 
    our circuit operates on a single noisy state, 
    reducing the circuit dimension and 
    thereby decreasing the likelihood of additional noise. 
    Furthermore, 
    the circuit can also be employed to probe the effects of AD noise. 
    We numerically calculated the fidelity
    between the purified and original states (channels), 
    as well as the corresponding success probabilities.

    Our results show that the proposed protocol 
    achieves a substantial enhancement in fidelity 
    while maintaining a relatively high success probability, 
    demonstrating its practicality and efficiency for experimental implementation.
    Since any quantum circuit can be decomposed into 
    a sequence of single- and two-qubit gates, 
    the purification of AD noise in single- and two-qubit systems 
    serves as a natural benchmark for 
    the design of large-scale noisy purification schemes. 
    Overall, these findings represent an important step 
    toward robust quantum information processing 
    and demonstrate the effectiveness of 
    noise-adapted purification strategies 
    in mitigating dominant decoherence mechanisms.
 

\bibliography{ref}
\end{document}